\documentclass{elsart}
\usepackage{natbib}
\usepackage{epsfig}
\usepackage{unites2e}
\begin{document}
\runauthor{LeBohec}
\begin{frontmatter}
\title{SGARFACE: A Novel Detector For Microsecond Gamma Ray Bursts}

\author[MCSD]{S. LeBohec}\author[UTAH]{}
\author[MCSD]{F. Krennrich}
\author[MCSD]{and G. Sleege}
\address[MCSD]{Physics and Astronomy Department, Iowa State University\\
               Ames, IA, 50011, USA}
 \address[UTAH]{Now at the Department of Physics, University of Utah\\
Salt-Lake-City, UT, 84112-0830, USA}     
\begin{abstract}
The Short GAmma Ray Front Air Cherenkov Experiment (SGARFACE) is  operated  
at the Whipple Observatory utilizing the Whipple 10~m $\gamma$-ray telescope.  
SGARFACE is sensitive to $\gamma$-ray bursts of more than $\rm 100~MeV$ with 
durations from $\rm 100$~ns to $\rm 35 \:\mu$s and  provides a fluence 
sensitivity as low as 0.8 $\gamma$ rays per $\rm m^2$ above $\rm 200$~MeV
(0.05 $\gamma$ rays per $\rm m^2$ above 2~GeV) and 
allows to record the burst time structure. 
\vspace{1pc}
\end{abstract}
\end{frontmatter}

\section{Introduction}

The detection of $\gamma$-ray burst phenomena at MeV-GeV energies is generally the
domain of satellite-based $\gamma$-ray telescopes with a wide field of 
view allowing sky surveys and monitors for classical Gamma Ray Bursts.  The 
sensitivity of these telescopes is limited by their collection area 
(0.1 - 1 $\rm m^2$ for EGRET, GLAST, respectively).
EGRET also had a dead-time  limiting the burst detection to time 
scales larger than 200~ms.  
Bursts of $\gamma$-rays with microsecond duration have been suggested to 
originate from primordial black holes \cite{hawking74}, assuming the 
final phase of their evaporation occurs rather fast, as suggested by some 
particle physics scenarios (for review see \cite{halzen91} and \cite{carr04}).
In this paper we present an exploratory ground-based $\gamma$-ray instrument 
that has a high fluence sensitivity to microsecond bursts of GeV $\gamma$-rays,
the SGARFACE experiment. 

Very High Energy (VHE) $\gamma$-ray astronomy has become 
an established discipline in
high energy astrophysics through the detection of more than 10  astrophysical
sources at TeV energies including supernova remnants, active galaxies,
the Galactic Center and a few unidentified sources (for review see 
\cite{weekes03}, \cite{hofmann04}). 
This recent development was made possible by the success of the imaging 
atmospheric Cherenkov technique pioneered by the Whipple collaboration using 
the Whipple 10~m $\gamma$-ray telescope on Mount Hopkins, in southern Arizona 
\cite{weekes89}.
Major next generation telescopes are currently under construction worldwide
and will provide an order of magnitude better sensitivity (see e.g., \cite{weekes02},
\cite{hofmann02}, \cite{enomoto02}, \cite{lorenz02}).

The success of the atmospheric imaging technique is based on its inherently
large effective collection area ($\rm 10^{4} - 10^{6} \: m^2$, depending on 
the $\gamma$-ray primary energy)  and  the ability to efficiently  
discriminate $\gamma$-ray showers from the much more numerous 
cosmic-ray  initiated  air showers by  off-line 
analysis of Cherenkov light images. A cosmic ray entering the atmosphere 
initiates a shower of secondary particles - mostly pions and their decay products 
including an electromagnetic component. The electromagnetic component of the 
cascade is accompanied by the formation of a Cherenkov light front which, 
at ground level has a duration of a few nanoseconds and extends  over more 
than $\rm \sim 130~m$ in radius for vertical showers. For example, a 
$\rm 1~TeV$ $\gamma$-ray primary results in a shower yielding a 
Cherenkov light density of  $\rm \sim 100 \: \: photons/m^{2}$ on the ground. 
When a telescope, equipped with fast  photo-detectors  and electronics, is 
located within the light pool it can be used  to record the Cherenkov flash, 
thus providing a large effective collection area. In case of a primary 
$\gamma$-ray, the image is elongated, pointing back to the direction of origin 
with a length of $\sim 0.5^\circ$ and a width of $\sim 0.2^\circ$
corresponding to the longitudinal and lateral spread of electrons and 
positrons in the shower.  
Cherenkov images from cosmic-ray showers are mostly broader, longer and 
the light 
distribution in the focal plane is more irregular and patchy than that 
from $\gamma$-ray induced showers.  Some cosmic-ray shower images 
also show Cherenkov rings and arcs from single muons passing close to the 
telescope.

The energy threshold of next generation  imaging atmospheric Cherenkov
telescopes will be 
lowered to explore the energy regime below 100~GeV. However,  for 
energies much below 100~GeV, the Cherenkov light density  becomes 
small  compared to the night-sky background, hence extremely large reflectors are
required.  It was suggested that telescopes with mirror  areas of order 1,000~$\rm m^2$ 
may be able  to reach a 5~GeV trigger thresholds using  existing technology 
\cite{aharonian02} if placed at 5~km altitude. 
However,  difficulties related to the shower physics arise from the fact that 
the Cherenkov light signal from a very low energy shower does carry rather 
limited information about the physical parameters (arrival direction, primary 
energy and impact point)  characterizing the primary $\gamma$-ray.   

Nevertheless,  $\gamma$-rays with energies as low as $\rm 100~MeV$ still 
produce small atmospheric showers with weak Cherenkov light yield. Such 
low energy atmospheric showers cannot be detected  individually through their 
Cherenkov emission. However, a large number of $\gamma$-rays 
arriving within a short period of time could produce a detectable glow of 
Cherenkov light.    Ground-based atmospheric Cherenkov telescopes could 
be highly sensitive to short duration  $\gamma$-ray bursts  with energies 
above $\rm 100~MeV$. This idea was first  suggested  by Porter and 
Weekes \cite{porter78}. They used a pair of telescopes separated 
by 400~km, each  
equipped with a single photo-detector searching for coincident light flashes 
from microsecond $\gamma$-ray bursts. Recently,  Krennrich, LeBohec and 
Weekes \cite{krennrich00} showed that short low energy $\gamma$-ray bursts 
could also be detected and identified with  a single imaging atmospheric 
Cherenkov telescope using FADCs and modern digital trigger electronics.  
Such a system can provide unprecedented sensitivity to $\gamma$-ray bursts 
in the range of time scales from $\rm 100~ns$ to $\rm 35~\mu s$, a regime 
still largely unexplored.

 This is the conceptual basis for the design of the Short GAmma Ray Front Air
Cherenkov Experiment (SGARFACE) which was installed on the Whipple 10~m 
telescope in Spring of 2003.   SGARFACE is the first experimental realization
of this idea using a modern Cherenkov telescope.  
 Here, we first review the properties of Cherenkov 
flashes that  should result from a $\gamma$-ray front entering the atmosphere. 
Then, we describe the design of the SGARFACE experiment and  present 
performance parameters to discuss its sensitivity to microsecond $\gamma$-ray 
bursts.

\section{The Cherenkov Light Glow From Multi-$\gamma$-Ray Initiated Showers}

Design and performance studies of a new air shower detection technique
requires the use of Monte Carlo simulations. For our studies presented in 
this paper we used an updated version of simulation programs that is based 
on ISUSIM (see \cite{mohanty98}).  After intense scrutiny and comparison 
between MOCCA and ISUSIM, these programs were both used to derive the 
Crab Nebula spectrum published by the Whipple collaboration \cite{hillas98}. 
The ISUSIM programs for the shower development and Cherenkov light production 
have been refined  in an updated program, known as GrISU. 
(http://www.physics.utah.edu/gammaray/GrISU). 
In GrISU, the timing of the shower development and light propagation 
has been introduced.

\subsection{Cherenkov Light Density From Individual Showers}
\begin{figure}[densite]
\epsfig{file=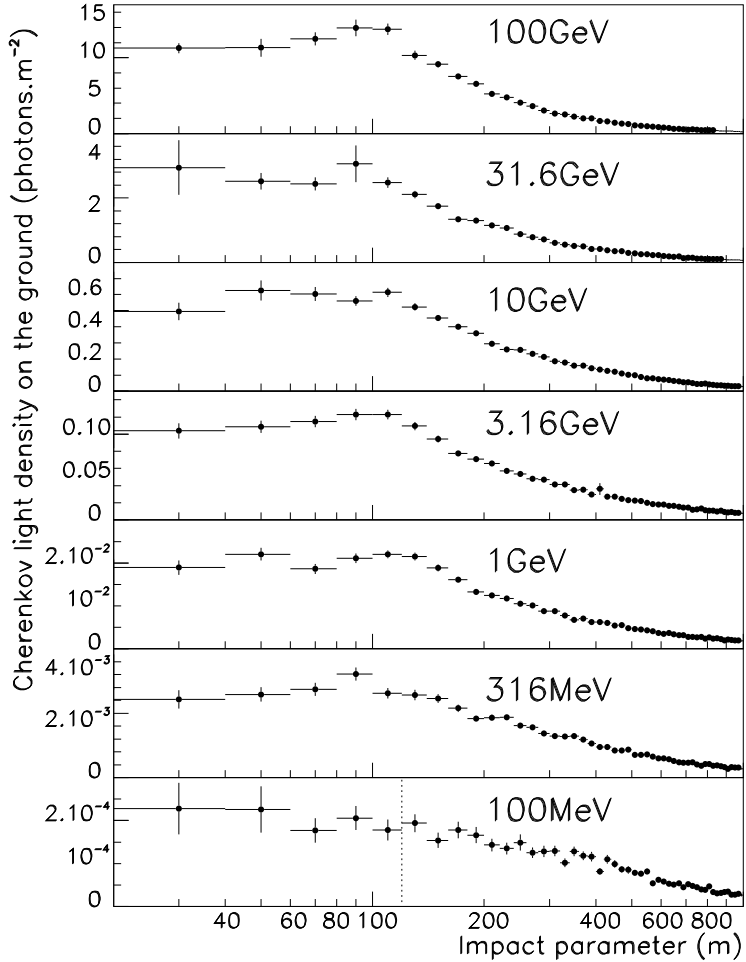,width=4in}
\vspace{10pt}
\caption{For vertical showers of primary energies ranging from 100~MeV to 
100~GeV, the Cherenkov light density at  ground level is plotted as a 
function of the primary impact parameter. }
\label{densite}
\end{figure}
Figure \ref{densite} shows the  simulated Cherenkov light density as a function
of impact parameter for the Whipple Observatory altitude (2,300~m above sea 
level) from individual $\gamma$-ray showers at zenith with primary energies 
ranging  from  100~MeV to 100~GeV with the US Standard Atmosphere absorption 
taken into account.
It can be seen from Figure \ref{densite} that the radial light distribution
exhibits a sharp rim at a distance of  $\rm \sim130$~m, which is most
pronounced for the highest primary energies (1~GeV - 100~GeV).  As described
in Hillas \cite{hillas90}, the increase of the Cherenkov angle, 
as particles penetrate deeper into the  atmosphere, is somewhat compensated for
by  the decreasing distance between the particle and the ground.  
This leads to an accumulation of  Cherenkov light  at  $\rm \sim 130~m$ shower core
distance, originating from a high energy particle component traveling along 
the shower axis \cite{bohec03}. 

Cherenkov light falling beyond this characteristic rim originates from  
particles that have experienced scattering at large angles.
Lower energy primary particles produce secondary particles that have a lower 
energy to start with, and hence the shower development tends to be more 
strongly affected by scattering. At energies below 1~GeV, this leads 
to a blur of the rim in the Cherenkov light density causing the lateral 
distribution to become flat out to distances of several hundreds of meters 
from the shower core.

\begin{figure}[size]
\epsfig{file=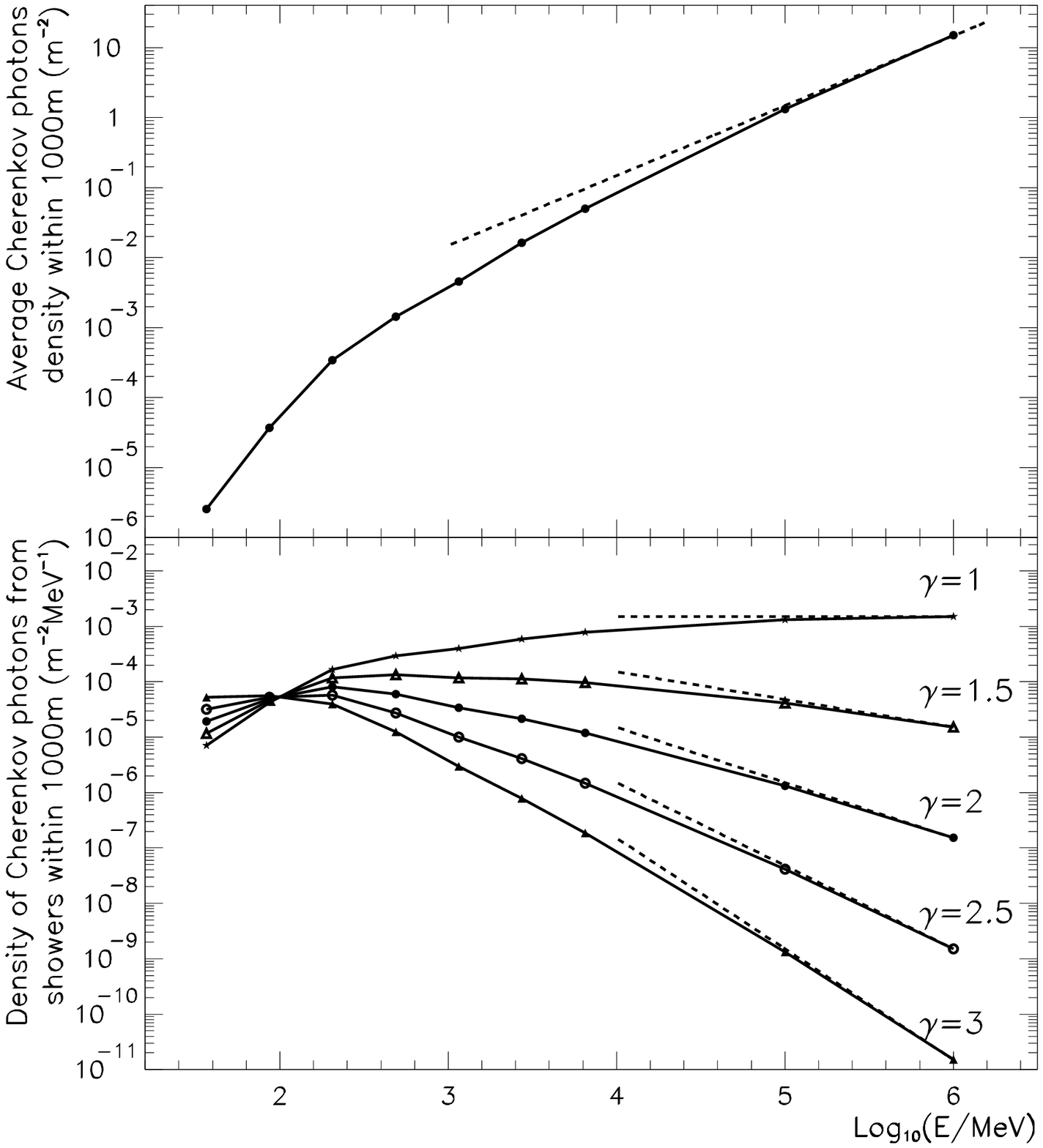,width=5.5in}
\vspace{10pt}
\caption{The top panel shows the average Cherenkov yield per square meter 
from vertical showers with impact parameters of less than 1,000~m as a 
function of the primary energy. The lower panel shows the same distribution 
multiplied by differential power law primary gamma-ray energy spectra of index 
1.0, 1.5, 2.0, 2.5 and  3.0 respectively, all with 1 $\gamma$-ray $m^{-2}$ 
above 100~MeV.}
\label{size}
\end{figure}

Figure~\ref{densite} also indicates that the Cherenkov light density  
from showers below 100~GeV does not scale linearly with the 
primary energy. This non-linearity has two origins. As the primary 
energy approaches the critical energy in the atmosphere, ionization  
losses become relatively more important and ultimately, $\gamma$-ray 
primaries of less than  $\rm \sim 100$~MeV do not develop an air shower 
and Cherenkov light emission ceases. Secondly, as the multiple 
scattering angle increases for lower energy secondary particles, the 
Cherenkov light is spread out over a larger area on the ground, reducing 
the Cherenkov light density.  

 The non-linearity of the Cherenkov light yield as a function of energy can be 
better seen in the top panel  of Figure \ref{size}, showing the average 
density of Cherenkov light on the ground within the arbitrary radius of 
$\rm 1000~m$. Figure~\ref{densite} shows that the contribution to the 
burst image from showers falling beyond 1000~m will be very small.
The linear relationship seen above 100~GeV is lost for lower energies and 
a cutoff occurs around 100~MeV. 

The lower panel of Figure~\ref{size} shows the Cherenkov 
light density per bin of energy produced by a $\gamma$-ray burst 
assuming various differential power law energy spectra 
($\rm dN/dE \propto E^{-\gamma}$) of index 1.0, 1.5, 2.0, 2.5 and 3.0, 
all with a fluence of 1~$\gamma$~ray per $\rm m^{-2}$ above 100~MeV.
 The effect of the 
cutoff can be seen below 100~MeV. At energies higher than 10~GeV, 
approaching the regime where the Cherenkov yield scales linear with the primary
energy, the response curve follows  a power law resulting from the chosen 
energy spectrum.
The peak response ranges from $\rm \sim 100~MeV$ for the $\gamma=3$ 
spectrum to 
arbitrarily higher energies for harder energy spectra. Astrophysical searches 
for which the SGARFACE experiment can be used have a wide range of possible 
energy spectra, from primordial black holes with thermal spectra at a few 
100~MeV to pulsed emission from $\gamma$-ray pulsars showing hard spectra 
($\rm dN/dE \propto E^{-1.4}  \: \: -  E^{-2.1}$ see \cite{nolan96}).

\subsection{The Cherenkov Light Image And Time Structure}

\begin{figure}[brstimg]
\epsfig{file=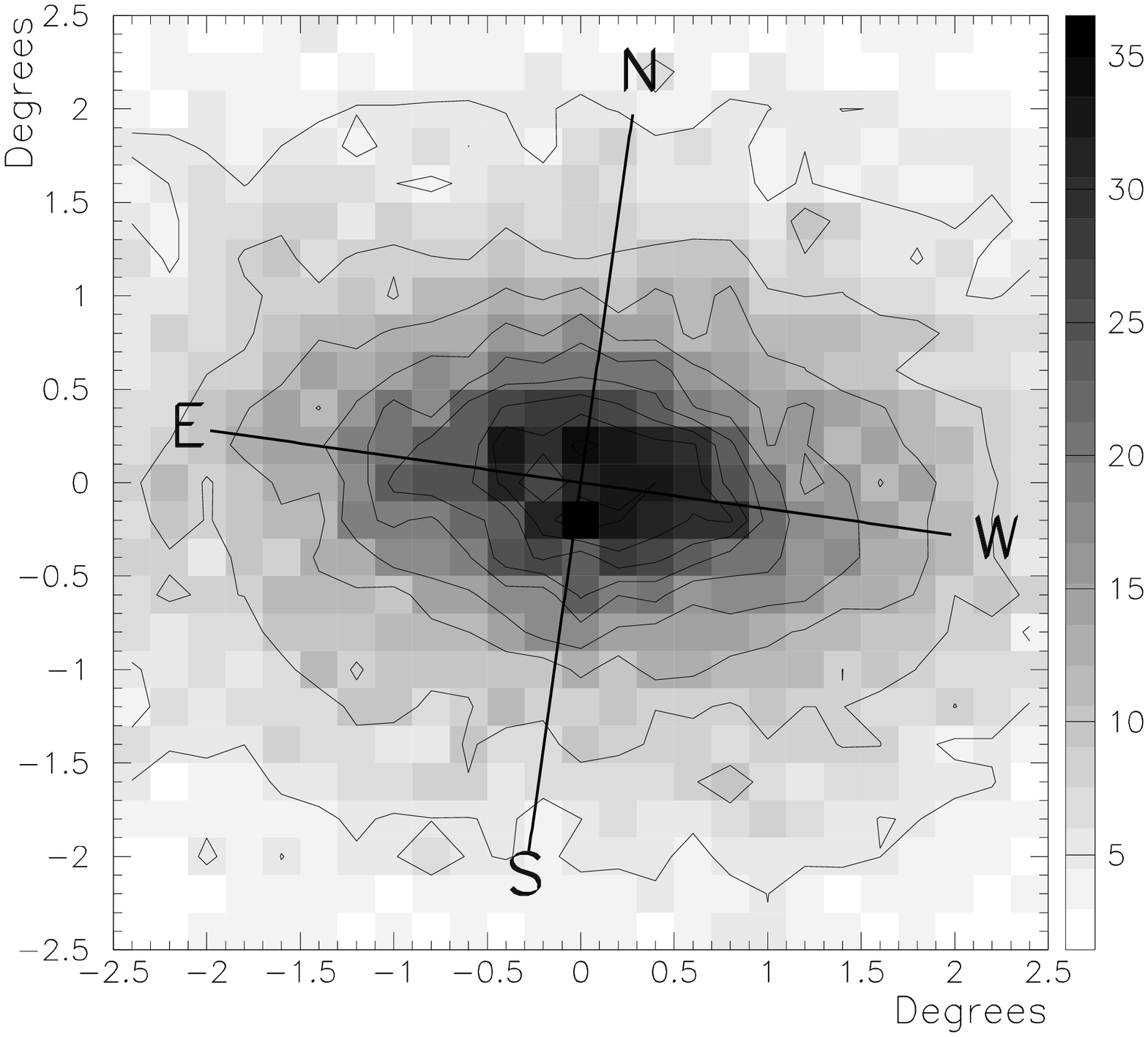,width=5.5in}
\caption{The Cherenkov light image produced by a $\gamma$-ray burst with a 
differential power law energy spectrum ($\rm dN/dE \propto E^{-\alpha}$) 
of index $\rm \alpha = $2.5 with fluence of 1.25 
$\gamma$-rays with  $\rm E >$~200~MeV per square meter. The gray scale 
indicates the number of Cherenkov photons per square meter and per angular 
bin of $0.2^\circ\times0.2^\circ$.}
\label{brstimg}
\end{figure}

Off-line analysis of Cherenkov light images has been very successful in
discriminating between $\gamma$-rays from discrete sources and cosmic-ray 
induced air showers. Both types of events 
potentially constitute a background to a search for short $\gamma$-ray bursts; 
however, the latter exhibit image and time properties which, when combined, 
render burst images unique.

Figure \ref{brstimg} shows the Cherenkov light image produced by a simulated
$\gamma$-ray burst at zenith with 
$\rm 1.25  \: \gamma~m^{-2}$ above 200~MeV. The differential energy spectrum 
taken is a power law of index 2.5 above 50~MeV.
 The cross indicates the direction of the burst and the orientation of the 
cross indicates the  direction of the Earth's magnetic field. The image is 
elongated in the direction perpendicular to the magnetic field. This effect 
results from the bent tracks of low energy electrons and positrons which 
allow showers with larger impact parameter in the east and west directions 
to contribute  Cherenkov light to the image. The magnitude of this 
effect depends on the angle between the magnetic field and the shower axis. 
It will be maximal in directions perpendicular to the magnetic field. 

%***********************************************%
% Magnetic field at MtHopkins from Mark Lang:   %
% Strength: 0.52Gauss                           %
% North: 11.5deg East                           %
% Dip Angle: 59.2deg                            %
%***********************************************%

%\subsection{The Cherenkov light time structure}
\begin{figure}[timedisp]
\epsfig{file=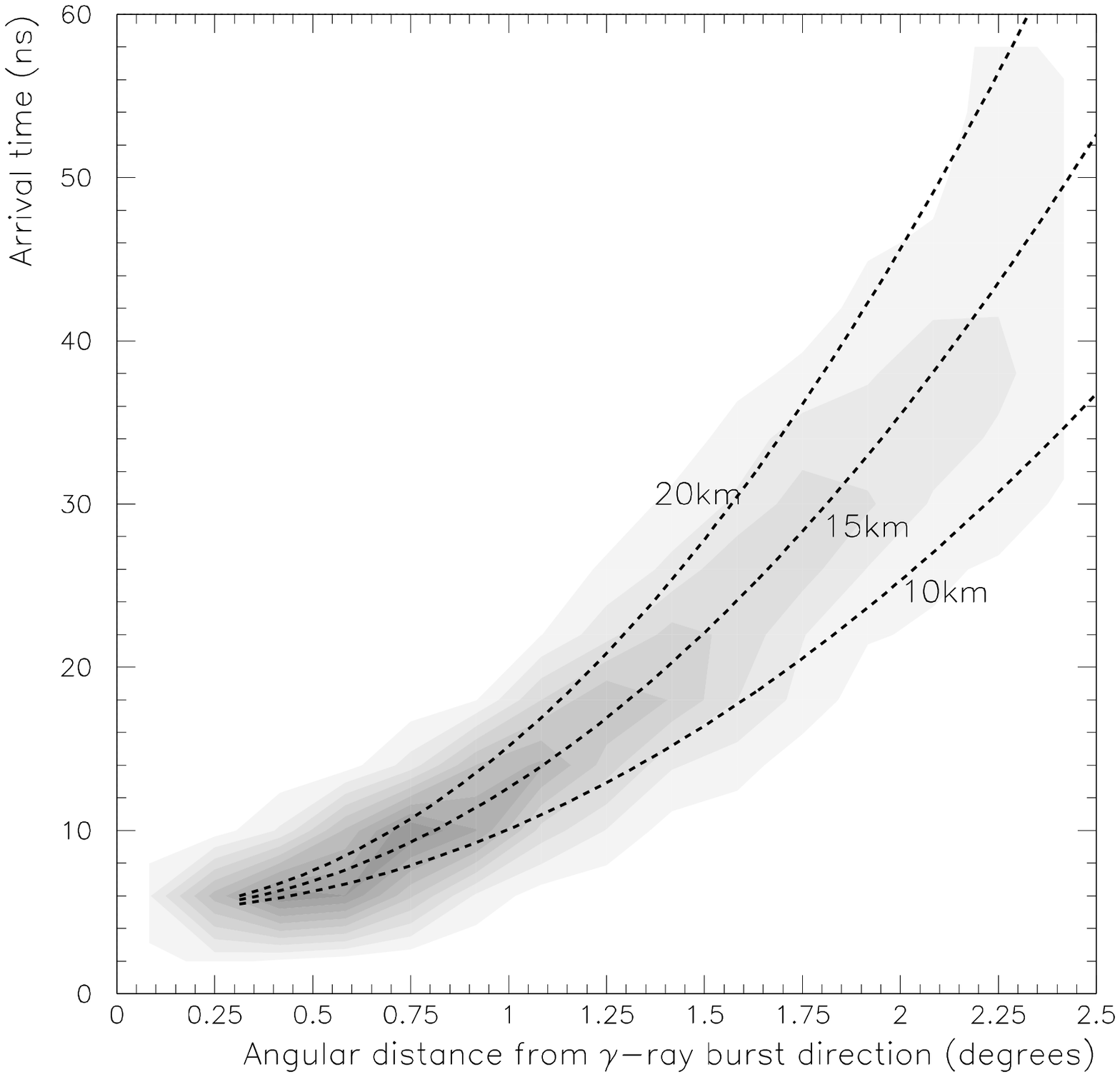,width=5.5in}
\vspace{10pt}
\caption{The arrival time of the Cherenkov light produced by a simulated 
$\gamma$-ray burst at zenith with a $\delta$-function time dispersion is shown 
as a function of the angular 
distance from the burst direction. The origin of time is the time at which 
the $\gamma$-ray front would have passed the detector if it had not interacted 
in the atmosphere. The dotted lines indicate the expected relationship if the 
emission occurred at altitudes of 10~km, 15~km and 20~km above the 
detector. }
\label{timedisp}
\end{figure}

The image time structure also contains potentially useful
information about the nature of the Cherenkov light image.
The angular distance of a Cherenkov photon from the burst direction is 
strongly correlated with the impact parameter of the  contributing low energy
shower.
 This is reflected in the time structure of the Cherenkov 
glow produced by a simulated $\gamma$-ray burst with no 
intrinsic time dispersion. The 
arrival time of the Cherenkov photons are shown in Figure \ref{timedisp} as 
a function of the angular distance from the burst direction. The relationship 
is well reproduced by the dotted lines which show the expected arrival time 
of the Cherenkov light emitted from fixed altitudes in a standard isothermal 
atmosphere.  The Cherenkov light in the outer part of the glow is coming from 
showers that are further away and have arrived later. The time dispersion 
ranges from $\rm \sim 5$~ns at the center of the image to $\rm \sim 10$~ns 
for $2^\circ$ away from the center. The time lag between those  points in the
image is $\rm \sim 30$~ns. As a consequence the time structure of the 
Cherenkov glow from a burst should be center symmetric and the signal in each 
point of the image should reproduce the burst time profile with a dispersion 
of less than 10~ns.

\subsection{Zenith Angle Dependence \label{zad}}

\begin{figure}[sensitpolar]
\epsfig{file=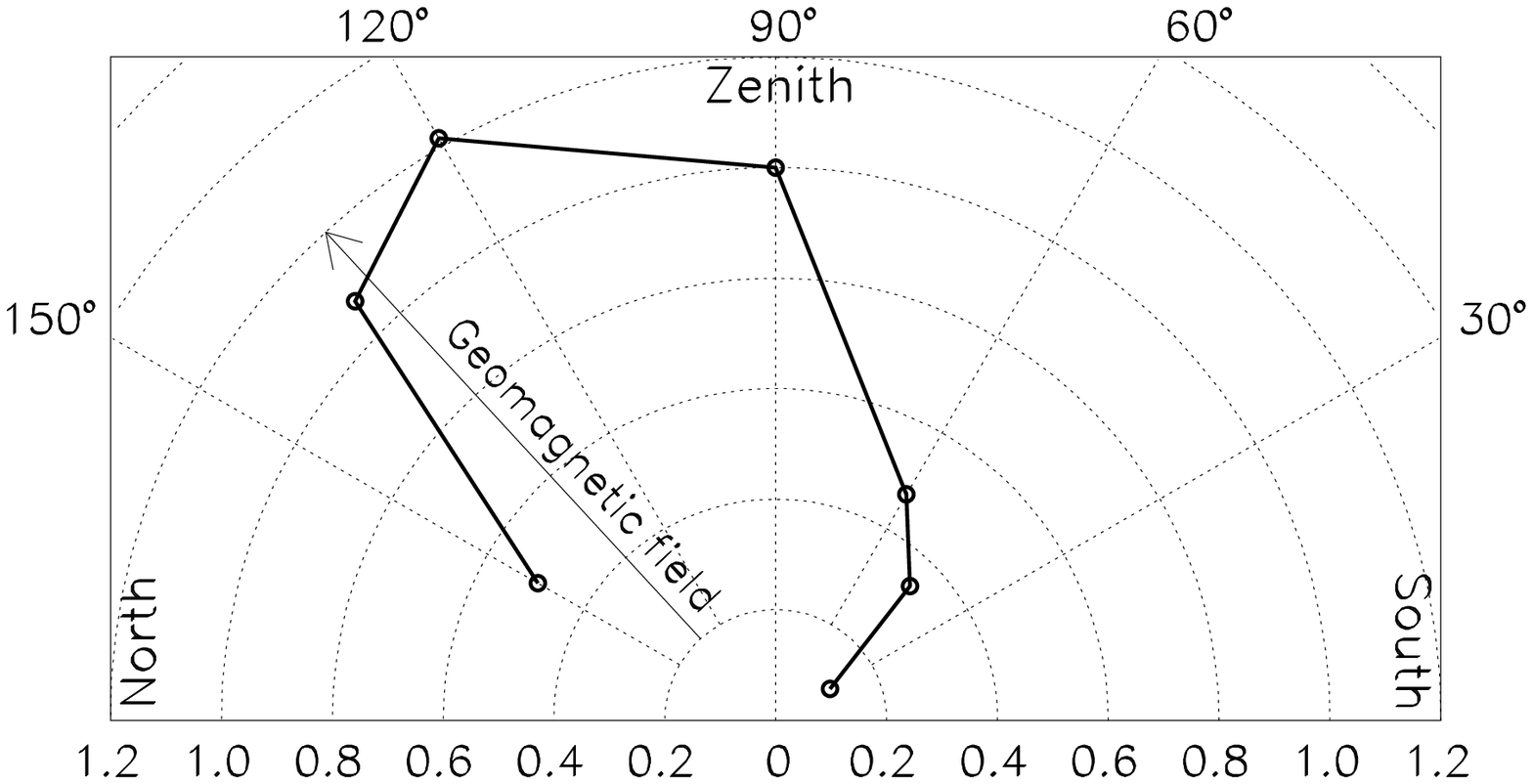,width=5.5in}
\vspace{10pt}
\caption{The geomagnetic field affects the light density in atmospheric 
Cherenkov $\gamma$-ray burst image. This effect is indicated on this polar diagram 
showing the relative sensitivity for bursts on the local meridian. The radial 
coordinate indicates the Cherenkov  light density in units of the light density at zenith.}
\label{sensitpolar}
\end{figure}

It is interesting to note that the burst detection technique is less dependent 
on zenith angle than the standard atmospheric Cherenkov technique detecting 
individual air showers; in the latter the energy threshold changes substantially 
with zenith angle. In the $\gamma$-ray burst detection, the energy  
threshold is fixed by the critical energy (typically 80 MeV) and does not 
depend on the zenith angle. Furthermore the different geometric effects 
compensate each other. As we go to larger zenith angles, electromagnetic 
showers develop further away from the telescope, the Cherenkov light is spread 
over a larger area and the fraction collected by the telescope decreases. 
This tendency is compensated by the fact that, for increasing zenith angle, 
each pixel collects the light from an increasing number of showers since 
showers develop at a larger distance. The sensitivity  to $\gamma$-ray 
bursts still somewhat decreases at lower elevation because of atmospheric 
absorption. 

A larger factor impacting the sensitivity of the burst technique is given by 
the geomagnetic field.  When  the angle between the line of sight and the  
Earth's magnetic field is maximal, the east-west extension of a burst image is  
increased while the Cherenkov light density of the image is reduced, thus reducing  
the sensitivity. When observations are made in the direction of the 
magnetic field the compactness of the image is at its  maximum and the 
sensitivity is not affected. This is illustrated in Figure \ref{sensitpolar} 
which shows the relative sensitivity along the local meridian. The lobe is 
skewed toward the celestial magnetic north pole; at $45^o$ elevation, the 
sensitivity is close to three times higher compared to the south at the same 
elevation.

\section{The Short GAmma Ray Front Air Cherenkov Experiment}

The properties of the Cherenkov light images produced by a $\gamma$-ray burst
are quite unique and we expect the possible detection of ultra-short 
$\gamma$-ray bursts to be essentially free of background 
events. In order to search for ultra-short
$\gamma$-ray bursts, we have designed and constructed an experiment 
to record Cherenkov light images and time profiles providing substantial 
sensitivity to microsecond scale $\gamma$-ray bursts. The Short GAmma Ray 
Front Air Cherenkov Experiment (SGARFACE) is designed to detect the glow of 
Cherenkov light collectively produced by atmospheric showers. 

\begin{figure}[experiment]
\epsfig{file=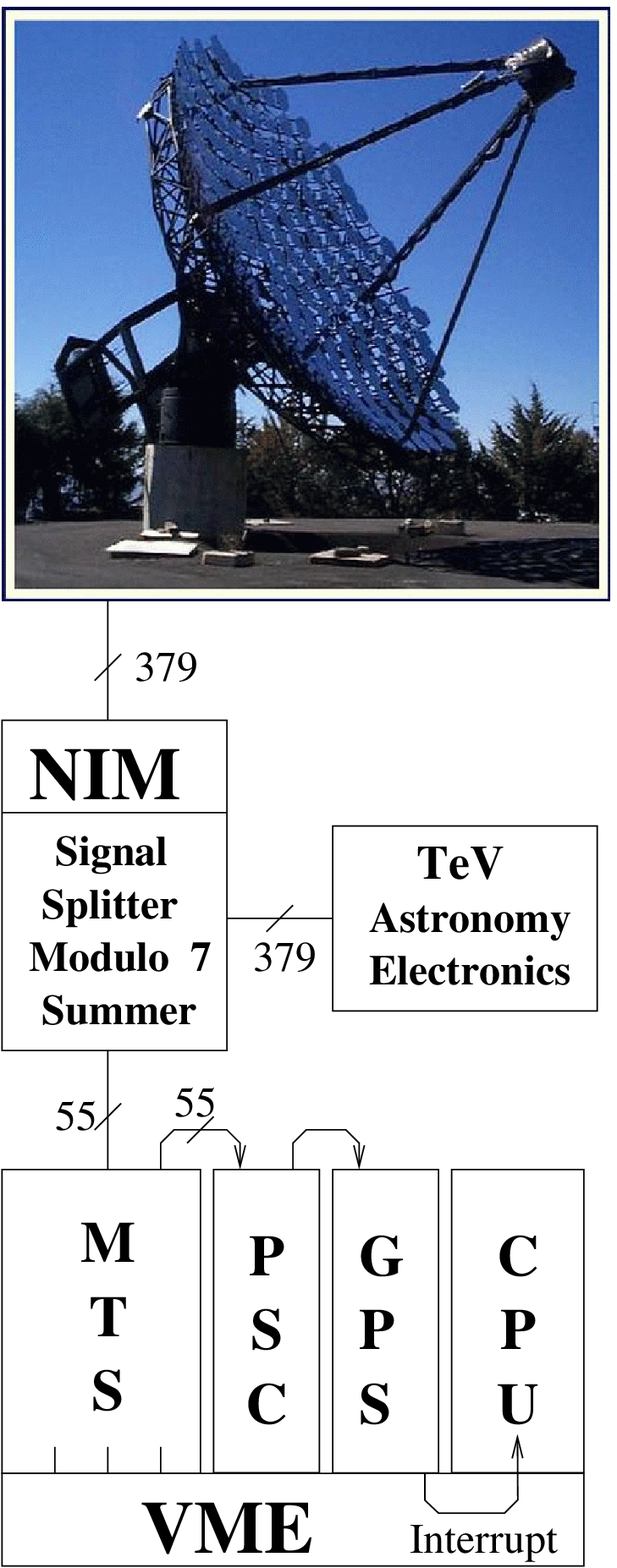,width=2.9in}
\vspace{10pt}
\caption{A schematic view of  experimental setup of SGARFACE within the Whipple
10~m $\gamma$-ray  telescope. Signals from the 379 photo-multipliers 
are duplicated before being sent to the Whipple standard electronics for TeV 
observations. The 55 SGARFACE signals are digitized and processed by the MTS 
discriminator modules. The PSC unit forms the trigger signal which is used to 
interrupt the system until all the information has been read by the computer.}
\label{experiment}
\end{figure}

SGARFACE was integrated in the Whipple 10~m  telescope in  Spring of 2003 
and it is operated in parallel to the standard TeV $\rm \gamma$-ray 
observations (see Figure \ref{experiment}).  
The Whipple 10~m reflector  has a mirror area of $\rm 70 \: m^2$.   The focal
plane instrument covers a field  of view  of $\rm 2.4^\circ$ with 379  
photo-multipliers  in a close  packed hexagonal  array with a $\rm 0.12^\circ$ 
center-to-center spacing \cite{finley01}. The photo-multipliers are operated 
at an average gain of  $1.1\times10^6$ \cite{krennrich01}.
%, a single photoelectrons corresponds to 3.3 digital count in the charge 
%integrating analog-to-digital converters of the Whipple standard electronics 
%system.
The  photomultiplier signals are sent via RG-58 cables (170 ft) to the 
nearby counting house where the electronics for VHE-$\gamma$-ray astronomy and SGARFACE is 
located. For further details on the Whipple 10~m telescope see \cite{finley01}.

In order to provide the photomultiplier signals to both the SGARFACE and
Whipple 10~m VHE experiments 
simultaneously, they are duplicated using a passive signal splitter 
(Figure \ref{experiment}).  At this level,  the signals  of 7 neighboring camera 
pixels are combined in an analog sum for further processing by the SGARFACE electronics. 

The 55 remaining analog signals are then  digitized by 50~MHz flash ADCs and 
buffered in the multi-time scale (MTS) discrimination  modules that constitute 
trigger level-1. The discriminator outputs are then used in the trigger level-2, 
a pattern sensitive coincidence unit (PSC) to reduce accidental rates from 
the night sky. When a level-2 trigger occurs,  the computer is prompted via 
hardware interrupt to read out the digitized pulse information buffered in the 
MTS. In this section, we describe the elements of the SGARFACE electronics 
system that were designed and constructed at the Electronics Design Center of 
the Physics Department at Iowa State University.

\subsection{Splitter-Summer Modules}
\begin{figure}[splitter]
\epsfig{file=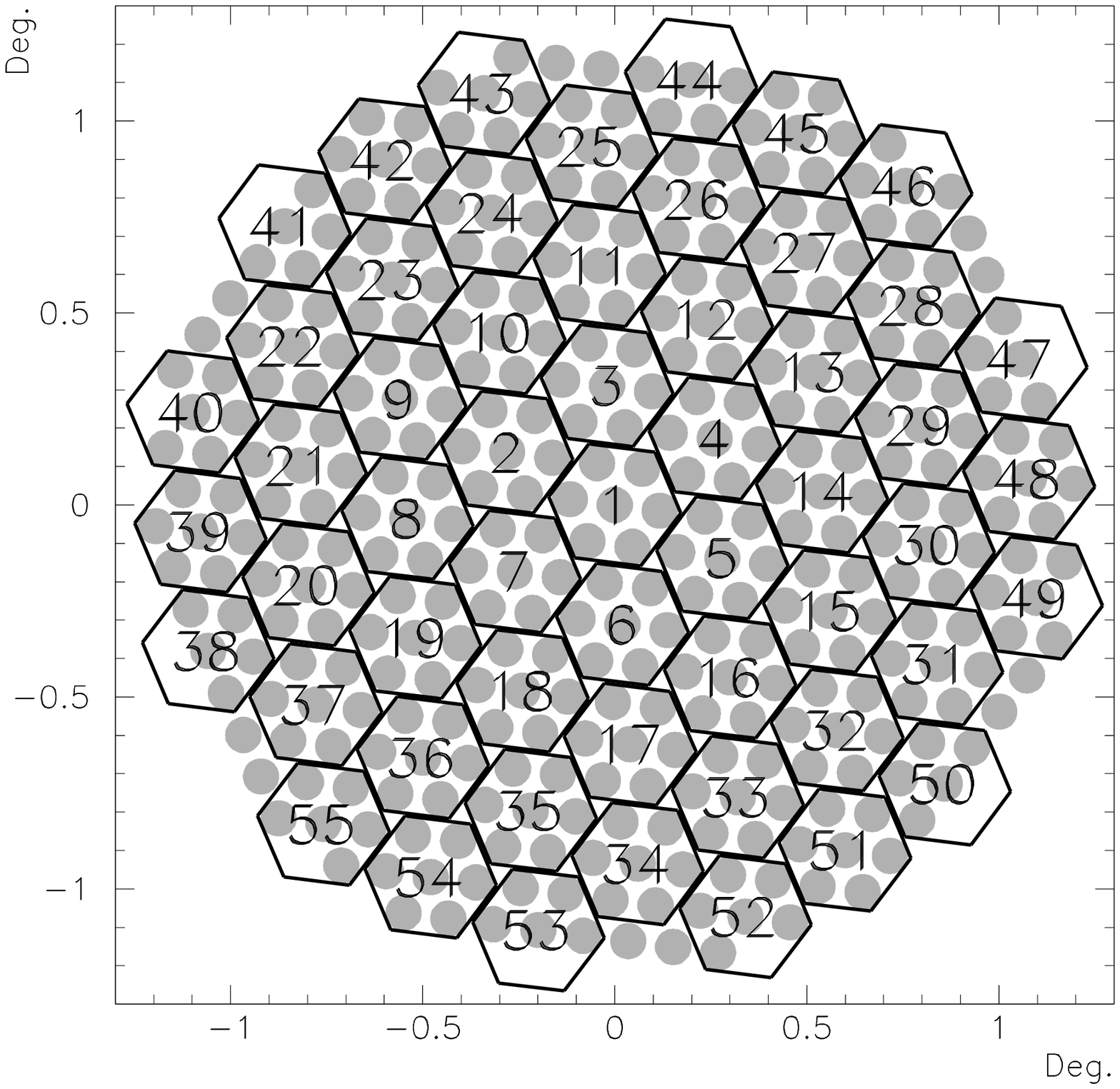,width=5.5in}
\vspace{10pt}
\caption{The re-mapping of the 379 pixels of the Whipple high resolution camera
into 55  clusters of 7 neighboring photomultiplier tubes into an analog sum. 
Each hexagon numbered from 1 to 55 corresponds to one SGARFACE channel.}
\label{splitter}
\end{figure}

The Cherenkov light image of a $\gamma$-ray burst is quite extended
($\rm \sim 1^\circ$) compared to the pixelation of the Whipple 10~m telescope 
camera ($\rm \sim 0.12^\circ$). This allows us to reduce the inherently high 
angular resolution of the Whipple camera for the integration of SGARFACE 
by combining the signals of 7  neighboring 
photo-multipliers (Figure \ref{splitter}).  The resulting pixelation  of  
$\rm \sim 0.4^\circ$ is adequate for triggering on $\gamma$-ray burst images. 

A passive splitter  preserves the signal bandwidth for the  standard Whipple
10~m VHE electronics introducing a $\rm 11\%$ attenuation.  A capacitive 
coupling before the summing 
junction ensures that the photomultiplier anode current component remains 
unaffected by the SGARFACE
electronics. It is used by the standard VHE system to monitor the night-sky 
luminosity. The rise time of photomultiplier signals  
can be as short as a few nanoseconds while the SGARFACE electronics 
digitizes the incoming signals in 20~ns slices. A bandwidth  limitation 
is applied to the SGARFACE signals by shaping them to a typical width of
more than 20~ns in order to minimize the information  loss during the 
digitization.  A voltage gain of 3 is applied to the signals going  to 
the SGARFACE digital trigger modules. 

The splitter and adder circuit boards are integrated in triple-width NIM
modules, with five boards per module. Each board provides inputs for 
seven photomultiplier signals entering through the back panel, and exiting 
through the front. The front panel also provides one dual output with the 
analog sum of 7 neighboring pixels. A total of 55 splitter-adder boards 
are used for the operation of SGARFACE.

\subsection{The Multi-Time-Scale (MTS) Discriminators}
Since SGARFACE is designed for use in a survey for $\gamma$-ray burst phenomena, 
the duration of a putative  burst is not known a-priori.   Therefore,  the
trigger has been designed to  cover the time scales for which this technique 
is most sensitive, a region largely unexplored by other instruments. 
The MTS discriminators, the level-1 trigger was designed to maximize 
sensitivity to pulses with durations from 100~ns to 35~$\mu$s. The 
incoming signals are amplified by an additional gain of 3 in the MTS 
discriminator module and then digitized by 50~MHz 8-bit FADCs covering a 
dynamic range of 1~V. Taking into account the photomultiplier gain, the 
attenuation in the cables and the various gains in the electronics, we find 
that 1 digital count corresponds to 1.8 photoelectrons per 20~ns sample.

The digitized signals are 
then used to form an integral over various fixed time intervals. This is 
achieved by summing the difference between the input and the output of a
first-in-first-out (FIFO) register stack through which the digitized 
signals are sent from the FADC. The number of registers in the stack defines  
the time interval over which the integral is calculated.  

\begin{figure}[mtsprincip]
\epsfig{file=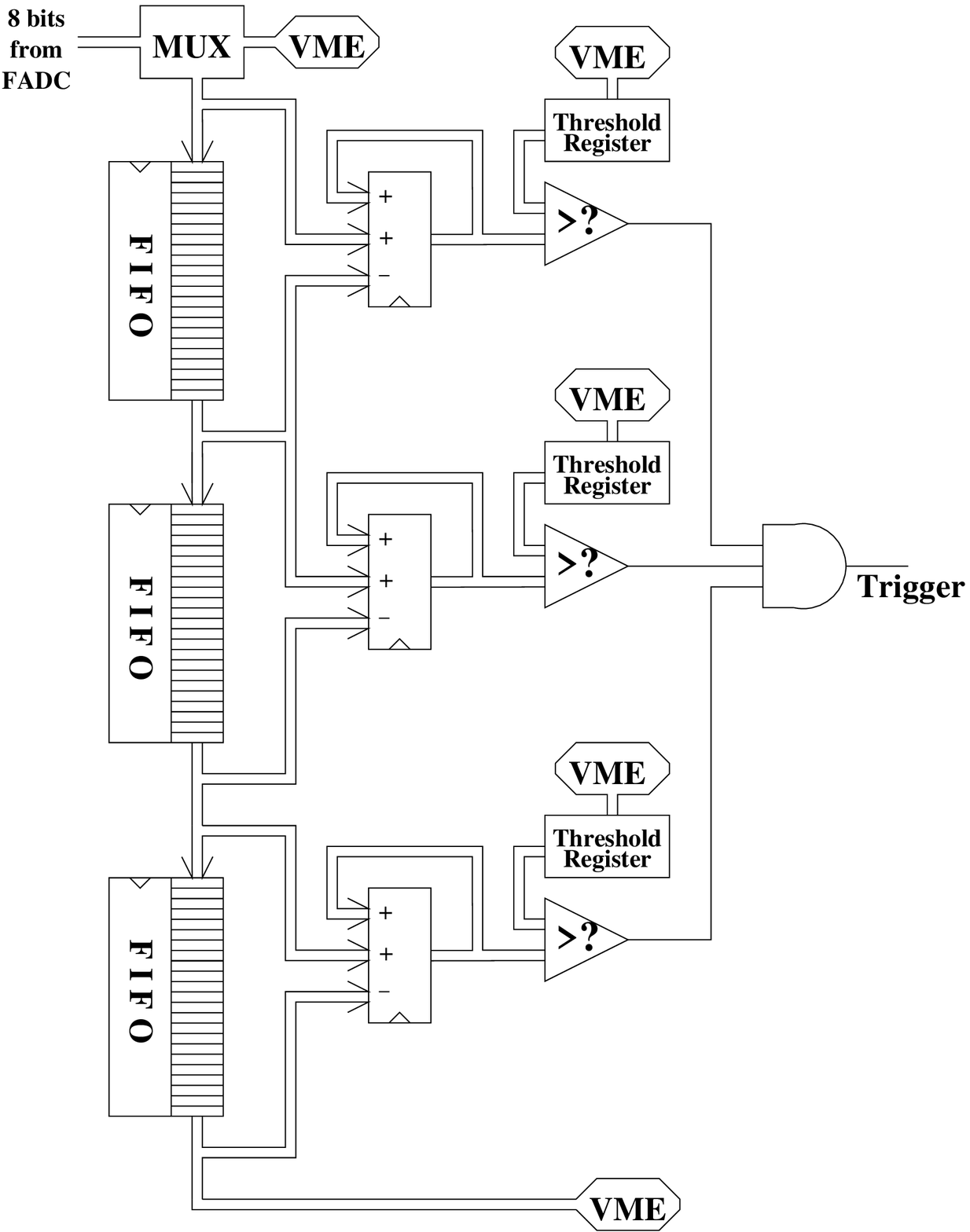,width=5.5in}
\vspace{10pt}
\caption{The signal integral over a time interval is obtained by 
accumulating the difference between the input and output values of a FIFO 
stack. The discriminator fires when the integrals over three consecutive 
time intervals exceed a predefined threshold at the same time.}
\label{mtsprincip}
\end{figure}

In fact, the signal is integrated over 3 consecutive time windows. A trigger 
signal is produced only when the three integral values exceed a predefined 
threshold at the same time (see Figure \ref{mtsprincip}). This  feature 
reduces the trigger sensitivity to 
signals that are shorter than  the time scale for which the trigger is 
designed. This is most relevant  for triggering on the shortest burst time 
scales of 60~ns. The 60~ns time scale trigger could still be sensitive 
to Cherenkov light pulses from cosmic-ray induced showers that are typically 
shorter than  40~ns \cite{krennrich00}. The requirement that at least  
3 consecutive time slices of 
20~ns show a signal, reduces the number of cosmic-ray triggers.

In order to provide sensitivity over a broad range of time scales, 
the logic using
FIFOs  is replicated in cascade. The first of the three time windows of each 
time scale is subdivided into three and used in the discrimination logic of 
the shorter time scale. This provides optimal sensitivity over time windows of 
60~ns, 180~ns, 540~ns, 1620~ns, 4860~ns and 14580~ns. The entire trigger logic 
is implemented in a $3^6=729$ register FIFO stack. In addition, the FIFO stack is 
followed by an extra 1024  register stack allowing us  to record  signal lengths 
up to  $\rm \sim 35 \: \mu s$.   

Moreover, each discriminator signal is  shaped to have a duration 
corresponding to its time scale. For each channel,  the six time scale 
discriminator signals are combined in an ``OR'' gate.  
The output of the ``OR'' gate is sent to the coincidence unit which can 
issue a global trigger signal (level-2) which is fed back into the MTS to 
freeze the register stacks. After a trigger 
occurred the local computer can read out the stream of data present in the 
MTS register stack.  The SGARFACE trigger system can then be reactivated.
The local computer can also write a diagnostic stream of data to any 
Xilinx FPGA to test for discriminator threshold and timing operations.

\begin{figure}[mtsboard]
\epsfig{file=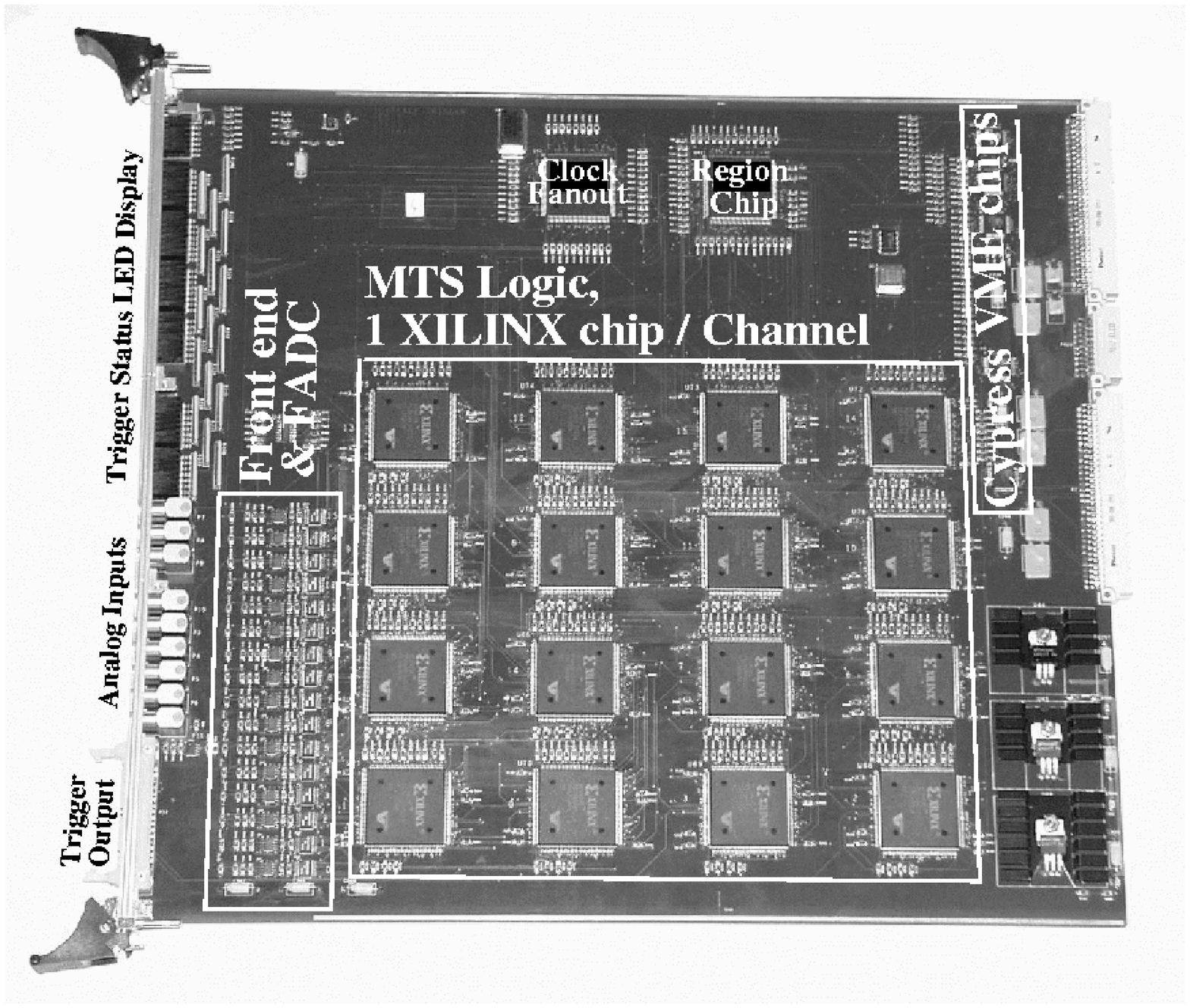,width=5.5in}
\vspace{10pt}
\caption{A 16 channel trigger level-1 board.  The 16 XILINX gate 
arrays in the center of the VME board contain the MTS discriminators and the 
register stacks. The gate arrays allow the formation of trigger decisions on 
time scales of 60~ns to 14.48~$\mu$s and the recording of Cherenkov pulses as 
long as 35~$\mu$s.}
\label{mtsboard}
\end{figure}

The  MTS discriminators are arranged in
 16-channel 9U VME-based modules (See Figure \ref{mtsboard}). Hence, four MTS 
discriminator modules are sufficient for the entire SGARFACE experiment.
 The VME interface is implemented around one  Actel FPGA permanently 
programmed as a region chip. It communicates with the 
VME back plane through four Cypress VME chips and their controller. This 
circuit handles commands, VME bus addressing and status reporting.
The MTS logic of each channel is implemented individually in a Xilinx 
XCV400E-PQ240 re-programmable gate array. All 16 Xilinx FPGAs are programmed 
by a local crate computer through the controlling region chip.  Instructions 
describing the trigger logic consist of a series of bytes written to the 
Xilinx FPGA consisting of both command and configuration data. After the 
Xilinx gate arrays have been programmed they can be addressed directly and 
information can be sent to and received from the local computer through 
simple ``{\it read}'' and ``{\it write}'' commands.  For example, each 
discriminator threshold value for individual channels and specific time 
scales is written to the Xilinx chip and can be read back by the local 
computer for verification. To test the performance of  the trigger level-1 it 
is possible to upload test data into the XILINX gate arrays, allowing a full 
verifications of the function of the digital trigger logic.

\subsection{Pattern Sensitive Coincidence (PSC) Unit}

Since the technique we describe here relies on triggering on extended Cherenkov
light images it is necessary to use a pattern selecting device that optimizes
the trigger efficiency for burst images. This is achieved by the PSC which 
detects coincidences within subsets of the 55 pixels across the camera. It 
permits to  require a programmable number of neighboring pixels to 
simultaneously deliver a level-1 trigger.

The PSC unit is designed to take 64 asynchronous inputs.   For the SGARFACE
experiment  a single PSC unit is sufficient for making a level-2 trigger 
decision based on the 55 inputs coming from the MTS unit. The design of the  
VME interface in the PSC unit is similar to that of the MTS 
discriminator. The PSC works as a  coincidence unit making a trigger decision
based on the topology of the event. 

In order to reduce the  accidentals rate from night-sky background 
fluctuations, the PSC can be made sensitive to up to 64 overlapping sectors. 
For each sector, the number of channels in a high state is calculated and 
compared to a multiplicity level required to issue a trigger. If any of the 
64 sectors reach a trigger decision,  a global trigger signal is issued. 
This signal is used to  hold the MTS discriminator modules, latch the GPS 
clock and  generate an  interrupt to notify the local computer that new 
data is available for readout. 
The logic is implemented in a single Xilinx XCV1000E-PQ240 FPGA chip.

\subsection{Software}
The VME control is based on a Tundra Universe II driver for Linux. The 
acquisition software is written in C. It launches an interrupt handler 
as well as a TCP/IP server. 

For each trigger event, the interrupt handler reads 
the time from the VME based GPS clock, the pulses stored in the MTS 
discriminator registers stacks and the trigger information from the PSC unit. 
The interrupt handler also reads the telescope tracking status as well as the 
photomultiplier high voltage values which are made available through NFS 
mounting of the disk of one of the computers of the standard 
Whipple 10~m electronics system.  
The event information is then written to disk and a log book is 
updated. The discriminator and the coincidence units which were stopped by 
the trigger are reactivated and the interrupt is re-enabled for the data 
acquisition to resume. 

The TCP/IP server accepts command strings which are 
interpreted and can act on the VME based modules by setting the various 
parameters of the experiment and send back information to the client for 
hand-shaking and display purposes. The TCP/IP client is a TCL/TK script which 
provides a graphical user interface for modifying the experiment 
configuration, start and stop the acquisition while offering a complete 
display of the status and most recent event.

\section{Performance}

\subsection{Trigger Rates}

\begin{figure}[biascurves]
\epsfig{file=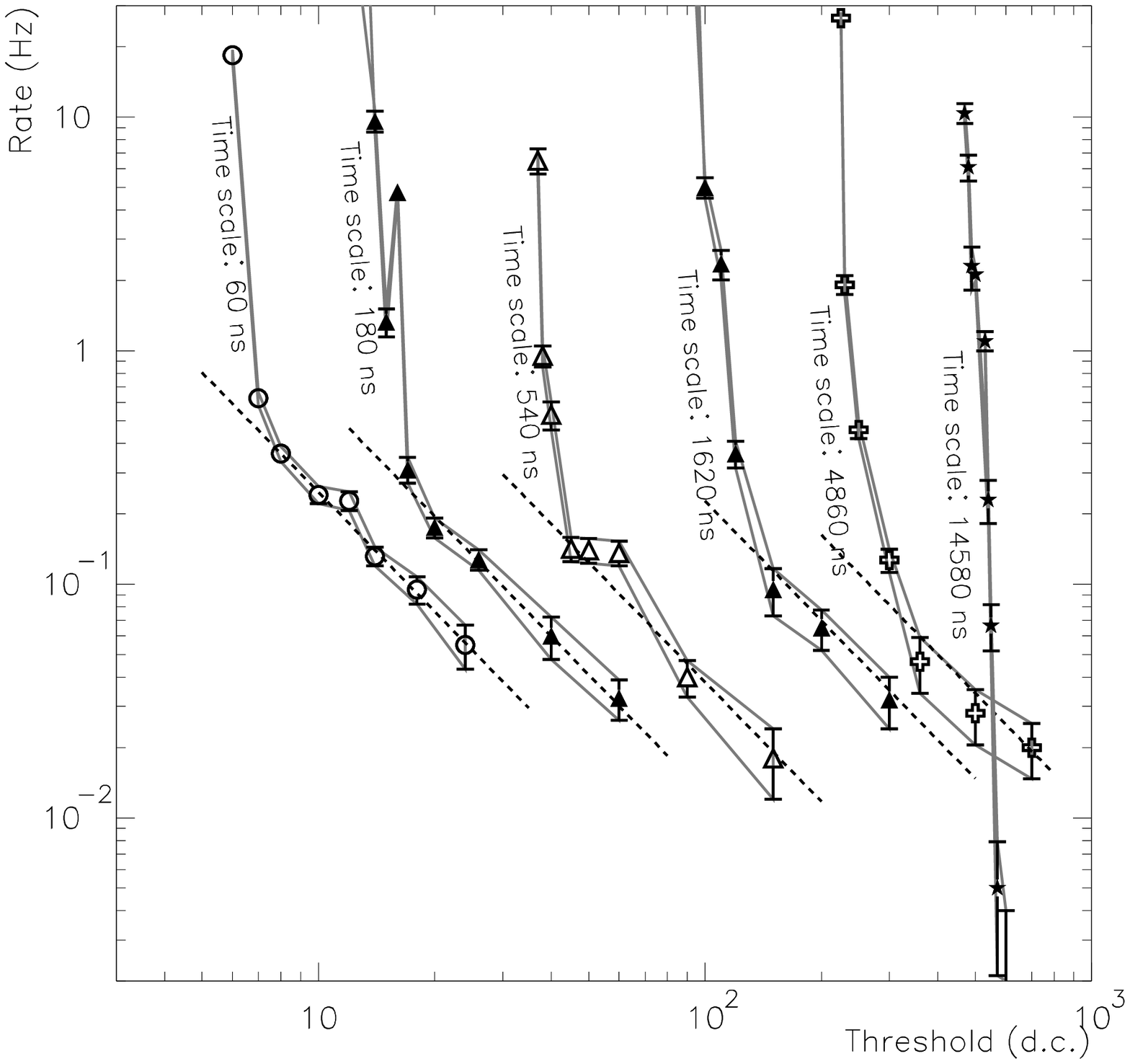,width=5.5in}
\vspace{10pt}
\caption{The trigger rate as a function of the threshold value for 
each of the six time scales is shown. For each curve, the steep slope corresponds to 
a noise fluctuation dominated triggering regime while, when the threshold is 
large enough, triggering results mostly from individual cosmic ray showers.}
\label{biascurves}
\end{figure}

We explored various trigger conditions with different multiplicities and
discriminator thresholds.   A suitable setting of the PSC unit is to
require a minimum  of 7 neighboring pixels exceeding the MTS discriminators. 
This gives 
sensitivity to Cherenkov light images that are about $\sim1.1^o$ across or 
larger.  The large multiplicity  is efficient in reducing accidentals and 
constitutes a good match to the angular extent of potential $\gamma$-ray burst 
images (\ref{brstimg}). 

In order to determine the optimal threshold settings 
of the MTS discriminators for each time scale, we measured the trigger rate
as a function of threshold (See Figure \ref{biascurves}). For small threshold 
values, the  trigger rate is dominated by night-sky background fluctuations 
and the rate decreases  rapidly with increasing threshold. For larger 
threshold values, triggers are predominantly due to  cosmic-ray showers. In 
this regime, the rate as a function of the threshold can be described by a 
power law. At the transition between these two regimes one can see a 
relatively sharp break in the slope of the trigger rate curve versus 
threshold (see Figure \ref{biascurves}). Generally, atmospheric Cherenkov 
experiments are  operated well above that break point to assure stable 
operation in the cosmic-ray dominated regime. 

All the different time scale triggers, except perhaps the shortest one, 
should not respond to a short Cherenkov flash from a cosmic-ray shower, which 
typically lasts less than 40~ns.  Nevertheless, all except the longest time 
scale do still show a power law tail corresponding to cosmic-ray triggers 
(Figure \ref{biascurves}). This results from an electronic artifact associated 
with the TeV electronics system: this system is not designed to preserve slow 
pulses, in fact it uses a high pass filter to reject pulses longer than 
200~ns. This causes a feedback in the splitter module, introducing a slow 
capacitive discharge pulse in the SGARFACE electronics. However, these events 
are easily rejected at the analysis  level since they show a short peak 
followed by a long tail.   

After determining  the breakpoint for each time scale, the MTS 
thresholds were set about  a factor of two above this transition  to avoid
instabilities in the trigger rates due to night sky variations. With these
stable settings the  event rate of SGARFACE is $\rm 0.3$~Hz with a typical 
dead-time of $\sim 10\%$ of the observation time. A single night of 
observations with 8~hours of on-time produces 800~Mb of data. As it turns out, 
these threshold settings for the different time scales do not scale as 
expected:  from Poisson noise fluctuations of the night sky, one 
would expect the threshold settings derived using trigger rate curves to 
increase with the square root of the  time scale. However, for the longest 
time scales the electronics noise  dominates over the night sky noise 
fluctuations. This is mainly due to the fact  that in the Whipple camera, 
the signal cables are only properly terminated  for the pulses passing the 
capacitive coupling. Slower components are  just reflected back and forth, 
affecting mostly the relatively slow pulses relevant for the SGARFACE 
experiment.  

Nevertheless, the sensitivity of SGARFACE is excellent. For the 180~ns time
scale, a trigger threshold of 48 d.c. (86 photoelectrons) accumulating over a 
time scale of 60~ns (one third of 180~ns) corresponds to 16 d.c. 
(28.8~photoelectrons) per 20~ns sample. For the $\rm 14.58 \: \mu$s time 
scale, a threshold of 1100 digital counts accumulating over $\rm 4.86 \: \mu$s
corresponds to $\sim 8.1$~photoelectrons per 20~ns sample. A SGARFACE channel 
is the analog sum of 7 neighboring photo-multipliers; therefore, for the second 
shortest and the longest time scales of SGARFACE the threshold corresponds to 
respectively $\sim 4.1$ and $\sim 1.1$ photoelectrons per photomultiplier per 
20~ns. To put this in perspective, the standard Whipple electronics has a 
trigger threshold of approximately 10 photoelectrons accumulating over a 20~ns
time interval in a single photomultiplier tube. Hence, depending on the time 
scale, SGARFACE with these conservative threshold settings is a factor of 2.4  
(180~ns) to 8.7 (14.58~$\mu$s) times more  sensitive to light densities than 
the standard TeV system.

\begin{figure}[event1]
\epsfig{file=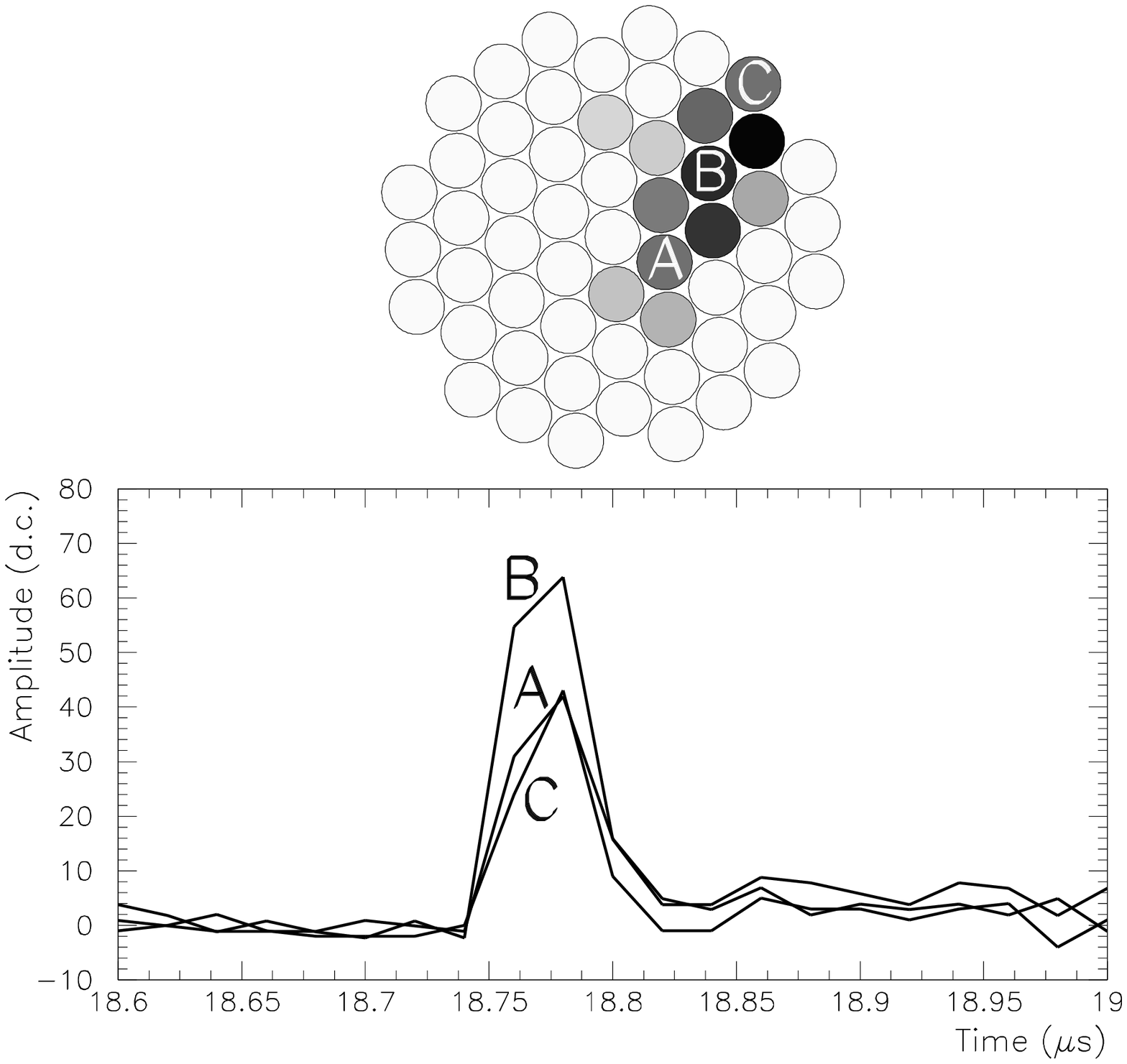,width=5.0in}
\vspace{10pt}
\caption{The Cherenkov light image from a cosmic-ray shower
and the pulse profiles from 3 channels are shown.}
\label{event1}
\end{figure}

\begin{figure}[event2]
\epsfig{file=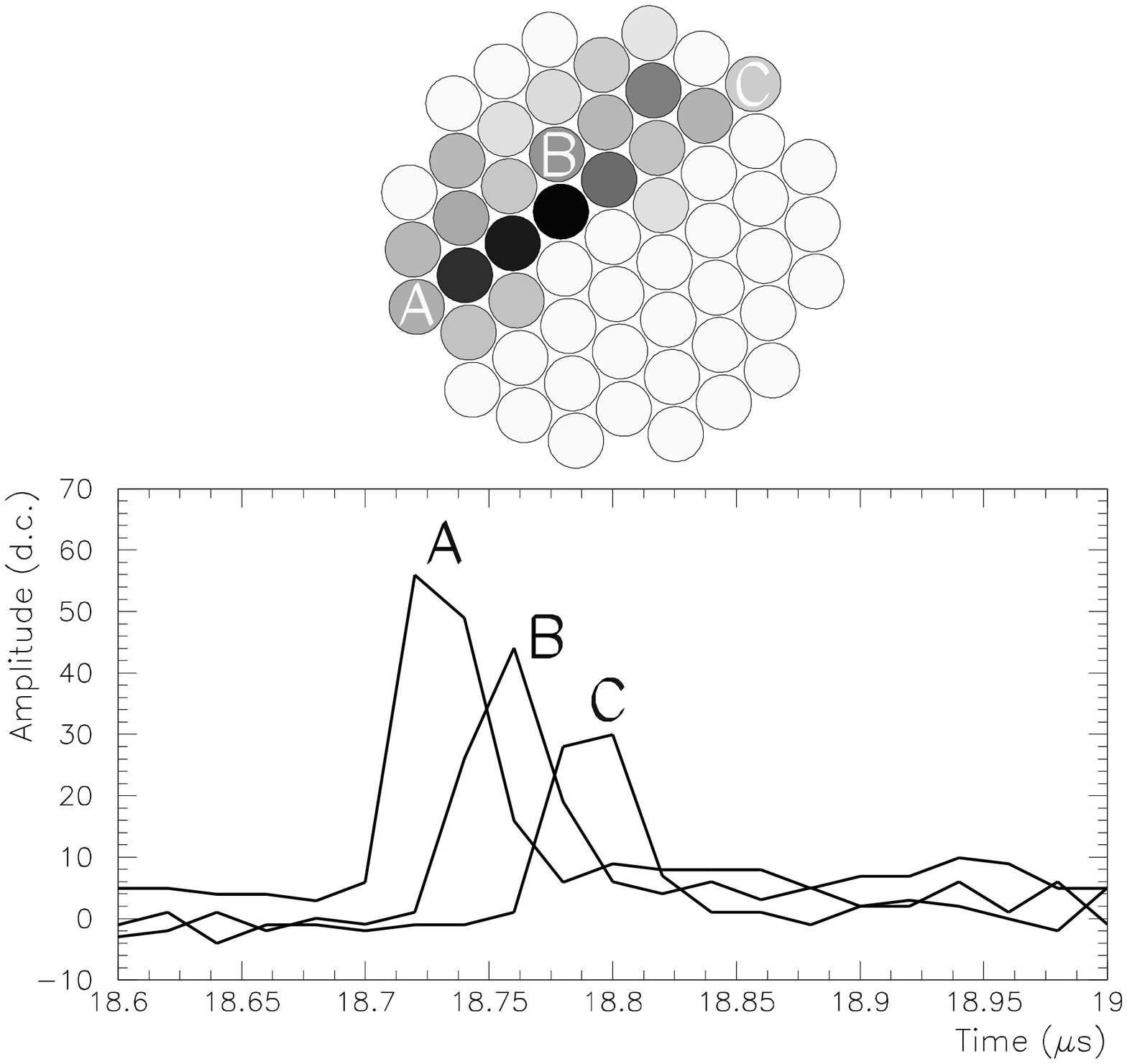,width=5.0in}
\vspace{10pt}
\caption{The Cherenkov light image from a cosmic-ray shower
and the pulse profiles of 3 selected channels indicating a time gradient 
across the image are shown. This suggests the shower arrival direction is largely 
off-axis.}
\label{event2}
\end{figure}

In Figure \ref{event1} we show the Cherenkov light image of a typical 
cosmic-ray event that triggered the SGARFACE electronics.  The image has been 
reconstructed using the flash-ADC information over a 60~ns time window.  The 
flash-ADC traces from three pixels line up within a few nanoseconds. 
Another event shown in Figure \ref{event2} shows a cosmic-ray image that 
extends all across the camera and the recorded pulses are systematically
shifted, indicating a time gradient along the shower image. This indicates the
shower axis was seen at an angle of several degrees, much larger than the
Cherenkov angle \cite{hess99}. This is possible since low energy particles 
are strongly deflected by multiple scattering resulting in scattering angles of several 
degrees from the shower axis. This increases the effective aperture of the 
camera making it possible to view events which are $\sim5$ degrees off-axis. 
Both events in figure \ref{event1} and \ref{event2} can be easily rejected 
as not resulting from a $\gamma$-ray burst because of their fast pulse and 
because of the time gradient along the image.

\subsection{Sensitivity}

In the previous section we established the hardware trigger threshold and
expressed the sensitivity of SGARFACE in terms of photoelectrons per time 
sample.  This information can be used to estimate the sensitivity to 
$\gamma$-ray bursts using simulations. A model of the  SGARFACE 
electronics  has been implemented 
in the detector simulation code of the GrISU package. The sensitivity of 
the experiment will depend of the actual burst energy spectrum and time 
profile. We have simulated $\gamma$-ray bursts with power law differential 
energy spectra of index 2.5 and a rectangular pulse profile of various  
widths. 

\begin{figure}[sensit]
\epsfig{file=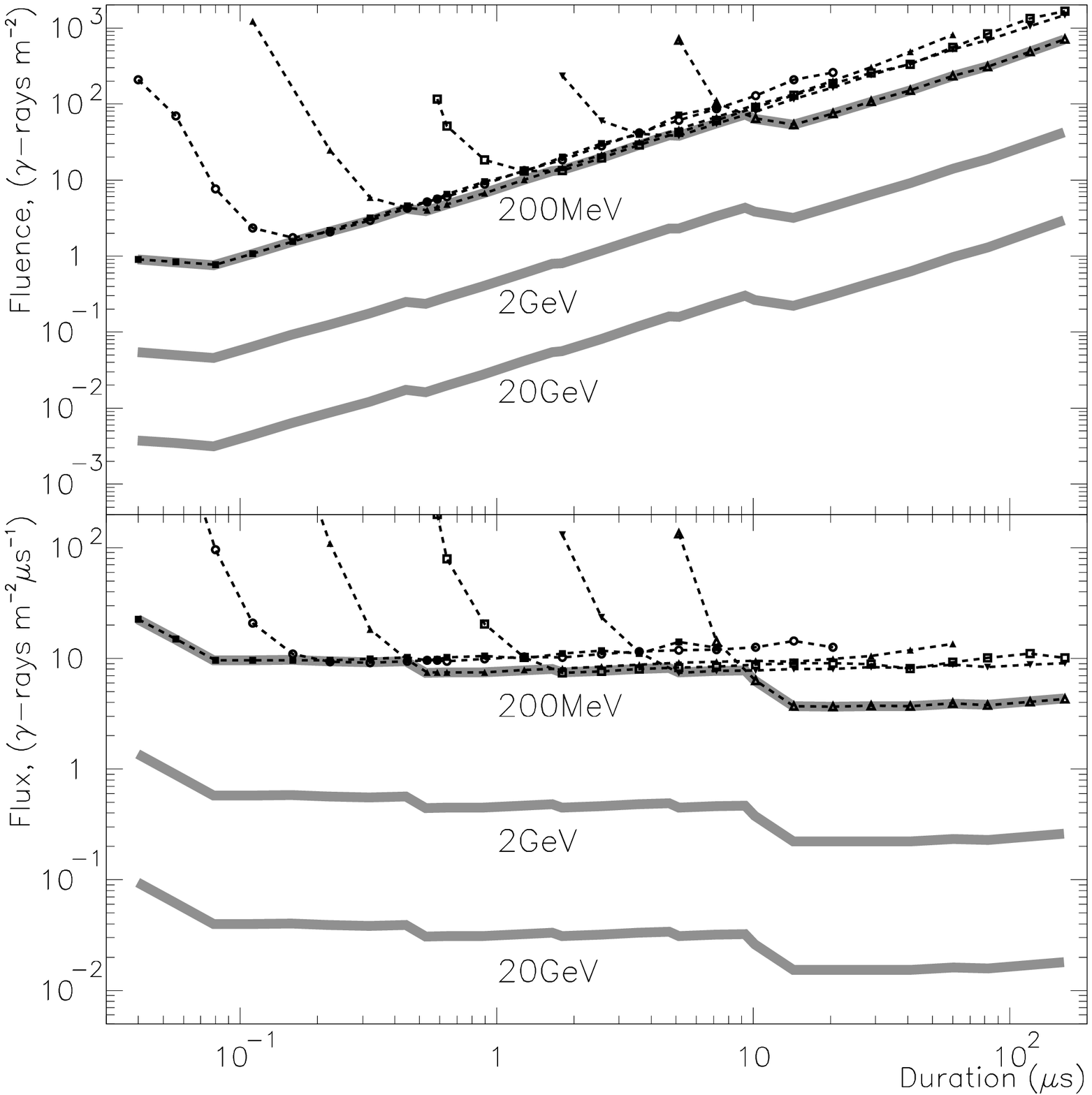,width=4.990in}
\vspace{10pt}
\caption{The sensitivity as a function of the burst duration in terms
of fluence (top) and in term of flux (bottom). The flux sensitivity curves 
better show the gain from the multi time scale sensitivity. The shaded lines 
show the combined sensitivity resulting from all six trigger time scales.}
\label{sensit}
\end{figure}

In Figure \ref{sensit} we show the fluence sensitivity to a $\gamma$-ray burst 
at zenith occurring on-axis, and therefore  centered within the field of view,   
as a function of  burst duration. Each curve corresponds to one of the 6 
trigger time scales ($\rm 60 \: ns$, $\rm 180 \: ns$, $\rm 540 \: ns$, 
$\rm 1.62 \: \mu s$, $\rm 4.86 \: \mu s$ or $\rm 14.58 \: \mu s$) showing the 
best sensitivity close to its specific integration time. For each time scale,  
the sensitivity to pulses with shorter durations is reduced as they 
are less likely to exceed the threshold simultaneously in three consecutive
integration time windows. For  pulses longer than the trigger time scale, a 
smaller fraction of the pulse is contained in the MTS discriminator 
integration window and the sensitivity is therefore also reduced. The 
sensitivity of SGARFACE corresponds to the envelope of the 6 curves as 
indicated by the broad grey line. With our current settings the SGARFACE 
sensitivity to bursts with differential energy spectra of index 2.5, ranges 
from  0.8 $\gamma$-rays $\rm m^{-2}$ to 50~$\gamma$-ray $\rm  m^{-2}$ above 
200~MeV for pulse width ranging from $\rm 60~ns$ to $\rm \sim 15 \: \mu s$, 
respectively.  However, the sensitivity above 2~GeV ranges from 
0.05 $\gamma$-rays $\rm m^{-2}$ to 3~$\gamma$-ray $\rm  m^{-2}$ for the
same time scales. These are sensitivities at zenith. As shown in section \ref{zad}, the sensitivity at 
other elevations will be affected mostly by the geomagnetic fields. Along 
the local meridian, the sensitivity improves at lower elevation to the north 
while it is rapidly degraded to the south as can be seen on figure 
\ref{sensitpolar}.

The trigger settings of the MTS units are very conservative with the
individual discriminators a factor of two higher than 
the break point from the trigger rate curves would indicate. With a next generation 
SGARFACE experiment attached to a state-to-the-art Cherenkov telescope
for which the noise would be dominated by the night-sky, the  sensitivity could 
be greatly improved, especially for the longer time scales.  

\begin{figure}[sensitheta]
\epsfig{file=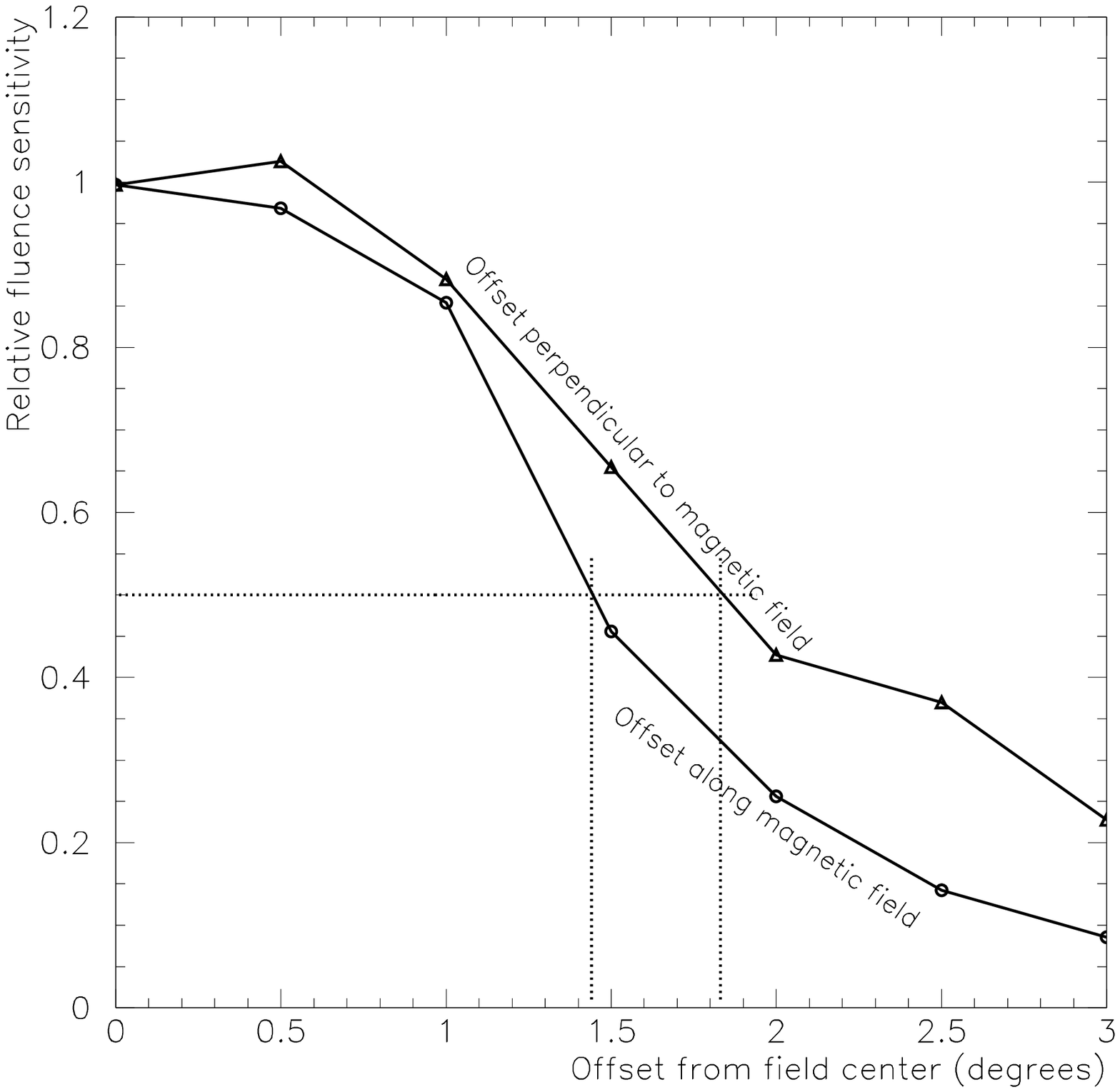,width=5.5in}
\vspace{10pt}
\caption{Because of the geomagnetic field, the sensitivity decreases
differently when the burst direction moves away from the field center along 
the magnetic field or orthogonally from it. This is illustrated here for a 
burst at zenith. The sensitivity is expressed in unit of sensitivity at field 
center.}
\label{sensitheta}
\end{figure}

Since the field of view of the Whipple camera is limited, it is important to 
estimate the sensitivity of bursts that arrive off-axis.   When the burst is  
off-axis, the sensitivity decrease does not depend on the time scale.
The relative sensitivity is shown in Figure \ref{sensitheta} as a 
function of the offset angle  for observations at  zenith. As a result of the 
elongation of the $\gamma$-ray burst Cherenkov 
image, the effective field of view also slightly depends on the arrival
direction with respect to the magnetic field:  it is larger for showers
arriving perpendicular  to the magnetic field. 
The fluence sensitivity remains constant within 10\% up to 
$1^\circ$ off-center and is reduced to 50\%  at $1.5^\circ$ off-center. 
 It should be noted that the instrument has sensitivity 
up to distances  of $\sim 3^\circ$. However,  when the burst image falls 
mostly outside of the field of view,  the event analysis and reconstruction 
will become less accurate. 

\section{Conclusions}

In this paper we provide a description of a ground-based $\rm \gamma$-ray 
detector with excellent fluence sensitivity to microsecond $\rm \gamma$-ray 
bursts at energies of 200~MeV and above. Gamma-Ray Bursts (GRBs) observed with 
space-based detectors show a wide range of durations - the exploration of the 
shortest time scales with these detectors is limited due to the detector 
integration time of typically more than 1~ms. Among the detected GRBs, 
millisecond and sub-millisecond variability is common \cite{walker00}. 
Because of this limitation of the space-based instruments to shorter time 
scales and their small effective collection area we have built an experiment 
that is complementary to space telescopes. We have designed and constructed a 
ground-based experiment (SAGARFACE) with sensitivity to microsecond bursts 
of $\gamma$-rays at energies above 200~MeV. The SGARFACE experiment has been 
operating since March of 2003 and in this paper we have presented the design 
and technical aspects of the apparatus. The essential part of the instrument 
is a flash-ADC system that is followed by XILINX re-programmable gate arrays 
which form the trigger on a wide range of time scales.   These electronics 
have been successfully implemented in the experiment and make the Whipple 
10~m telescope sensitive to microsecond bursts of $\gamma$-rays.

SGARFACE has an unprecedented fluence sensitivity  of 0.8~$\gamma$-rays 
$\rm m^{-2}$ for bursts lasting 0.1~$\mu$s and a fluence sensitivity of 
6~$\gamma$-rays $\rm m^{-2}$ for microsecond bursts at energies above 200~MeV.
When comparing with  EGRET's fluence sensitivity for short bursts,  a regime  
for which the EGRET instrument would require the  detection of multiple photons
within a trigger gate of 600 ns, we see that SGARFACE has a factor of 100 
(20) better fluence sensitivity at 100~ns (1$\mu$s). For bursts with spectra
peaking at 2~GeV or above, the fluence sensitivity is a factor of 2500 better 
than EGRET. As bursts are not expected to be more frequent in some directions 
than others, SGARFACE data are taken toward whichever target the Whipple 10m 
telescope is tracking and data is taken in the galactic plane as much as away 
from the galactic plane. Data taken in the direction of the Crab Nebula will 
be used in a dedicated analysis to search for counterparts to pulsar giant 
radio pulses \cite{shearer}.

A limitation of the SGARFACE experiment is its small field of view (3~degrees
across). Nevertheless, a survey using approximately 1000~hours of observations 
is expected to have a sensitivity that is substantially better than previous 
experiments for detecting primordial black holes  in the microsecond time 
scale regime. If SGARFACE were extended to an array of telescopes it would 
truly become a background-less experiment as for a $\gamma$-ray burst event, 
no parallactic displacement of the image is expected from telescope to 
telescope. This peculiar property would make it virtually impossible to 
misclassify an event as a $\gamma$-ray burst. The detection of a few 
interesting events with the current SGARFACE would make a strong case for an 
extension to VERITAS. With next generation wide field of view atmospheric 
Cherenkov detectors a SGARFACE trigger could be used as a very effective 
survey for ultrashort GRBs. 

\section{Acknowledgment}

We thank Harold Skank for help in the electronics design during the early
development stage of the project. We also thank  Roy McKay for invaluable 
technical assistance in the assembly and test of the prototype trigger boards. 
We acknowledge Oscar Valfredini and Louis Le~Bohec for their help in the 
cabling of the experiment. FK acknowledges the support for the SGARFACE 
project by the Department of Energy High Energy Physics Division through the 
Outstanding Junior Investigator program and generous financial support by Iowa 
State  University.  Also we would like to express many thanks to Trevor 
Weekes for reading the manuscript.  We  thank the VERITAS collaboration for 
the usage of the Whipple 10~m telescope.

\end{document}